\documentclass[12pt]{article}  

 \usepackage{cite} 
\usepackage{amssymb} 
\usepackage{amsfonts} 
\usepackage{amsmath} 

\usepackage{breqn}
\usepackage{slashed}

\normalbaselines
\catcode `\@=11
 
\def\section{\@startsection {section}{1}{\z@}{-3.5ex plus -1ex minus
     -.2ex}{2.3ex plus .2ex}{\normalsize\bf}}
\def\subsection{\@startsection{subsection}{2}{\z@}{-3.25ex plus -1ex minus
 -.2ex}{1.5ex plus .2ex}{\normalsize\bf}}

\def\thebibliography#1{\section*{References\markboth
  {REFERENCES}{REFERENCES}}\list
  {[\arabic{enumi}]}{\settowidth\labelwidth{[#1]}\leftmargin\labelwidth
  \advance\leftmargin\labelsep
  \usecounter{enumi}}
  \def\newblock{\hskip .11em plus .33em minus -.07em}
  \sloppy
  \sfcode`\.=1000\relax}
 
\catcode `\@=12 

\title{\bf\normalsize 
ON PRECANONICAL QUANTIZATION OF GRAVITY \\ IN SPIN CONNECTION VARIABLES
 }

\author{ 
Igor V. Kanatchikov$^{}$\thanks{{\sl E-mail address}: 
{\tt kanattsi@gmail.com, kai@fuw.edu.pl}} \\
\small 
National Center of Quantum Information in Gdansk 
(KCIK),\\ 
\small 81-824 Sopot, Poland 
} 

\date{\sf\small 
} 

\newcommand{\beq}{\begin{equation}}
\newcommand{\eeq}{\end{equation}}
\newcommand{\beqa}{\begin{eqnarray}}
\newcommand{\eeqa}{\end{eqnarray}}

\newcommand{\pbr}[2]{ \{ \hspace*{-2.6pt} [ #1 , #2\hspace*{1.4 pt} ] 
\hspace*{-2.6pt} \} }

\newcommand{\we}{\wedge}
\newcommand{\der}{\partial}

\newcommand{\inn}{\hspace*{2pt}\raisebox{-1pt}{\rule{6pt}{.3pt}\hspace*
{0pt}\rule{.3pt}{8pt}\hspace*{3pt}}}

\newcommand{\ga}{\gamma}

\newcommand{\ka}{\varkappa}

\newcommand{\Psib}{\overline{\Psi}}
\newcommand{\Phib}{\overline{\Phi}}

\newcommand{\what}[1]{\widehat{#1}}

\newcommand{\bx}{{\mathbf{x}}}

\newcommand{\BPsi}{\bf \Psi} 
   
\begin{document}

\maketitle

  


\begin{abstract}
\noindent
 The basics of precanonical quantization and its relation to the 
 functional Schr\"od\-inger picture in QFT are briefly outlined.   
 The approach is applied to quantization of Einstein's gravity 
 in vielbein and spin connection variables and leads to 
 a 
 quantum dynamics described by the covariant Schr\"odinger equation for 
 the transition amplitudes on the bundle of spin connection coefficients 
 over the space-time, 
 that 
 yields a novel quantum description of space-time geometry. 
 A toy model of precanonical quantum cosmology based on 
 the example of flat FLRW universe is considered.
 
\vspace*{0.5cm}



\vspace*{0.5cm}
\end{abstract}

\section{Introduction} 
We are accustomed to the notion that fields are infinite dimensional Hamiltonian systems and 
the fact that time plays a special role in the formalism and interpretation of quantum theory. Both aspects can be seen as inherited from the canonical Hamiltonian treatment which is a basis of canonical quantization  both in quantum mechanics and quantum field theory.   
However, it is much less known that the Hamiltonian formulation can be extended to field theory without explicitly distinguishing the time dimension and without 
referring to the infinite dimensional configuration or phase space. 
The examples of such an extension are  known in the calculus of variations of multiple integrals,  
    the simplest of them is the De Donder-Weyl (DW) theory 
(see e.g. \cite{dw}). 

The DW Hamiltonization of a field theory given by the first order Lagrangian 
$ L = L(y^a, y^a_\mu, x^\nu) $ is based on the covariant Legendre transformation 
to the new set of variables: { polymomenta } 
$$p^\mu_a := \frac{\der L}{\der y^a_\mu}$$ 
and the {DW  Hamiltonian function } 
$$ H(y^a, p^\mu_a, x^\mu) := y^a_\mu(p) p^\mu_a - L.$$
If the transformation is regular, i.e.  
$$\det \left |\left | {\der^2 L}/ {\der y_a^\mu\der y_b^\nu }
\right |\right| \neq 0 ,$$ 
the field equations can be cast into the 
{DW covariant Hamiltonian form:} 
\beq \label{dw}
\der_\mu y^a (x) = \frac{\der H}{\der p^\mu_a} , \quad 
\der_\mu p^\mu_a (x) = - \frac{\der H}{\der y^a}  . 
\eeq

Precanonical quantization 
is aimed at the construction of quantum field theory based 
on the space-time symmetric version of the Hamiltonian formalism like the DW formulation. It has been demonstrated 
(see e.g. \cite{garcia}) 
that the mathematical structures of the DW formulation
(also known as multi- or polysymplectic)  are in a sense intermediate 
between the Lagrangian and the canonical Hamiltonian description, whence the 
term "precanonical" proposed by us.

In the previous papers \cite{myprecanonical}, to which I refer for more details,  
it was argued that precanonical quantization leads to the following 
representation of the operators of polymomenta: 
\beq
\hat{p}{}^\nu_a = -i \hbar \ka \gamma^\nu 
\frac{\der}{ \der y^a},
\eeq
which act on the Clifford-valued (in general) wave functions $\Psi (y,x)$ on the 
finite dimensional space of field and space-time variables $y$ and $x$. 
The constant $\ka$ naturally appears in precanonical quantization 
on the dimensional grounds:   its  dimension in $n$ space-time dimensions  
is ${\tt cm}^{\tt 1-n}$ 
and it is related to the inverse of a very small "elementary volume" 
 as it appears e.g. in the representation of the $(n-1)$-forms  
 $\varpi_\nu:=\der_\nu \inn (dx^0\we...\we dx^{n-1})$ 
in terms of the Clifford algebra elements:  
 $\what{\varpi_\nu} = \frac{1}{\ka}\gamma_\nu$. 
 
The covariant analogue of the Schr\"odinger equation 
in precanonical quantization takes the form: 
\beq \label{nse}
i\hbar\ka \gamma^\mu\der_\mu \Psi = \what{H}\Psi, 
\eeq
where $\hat{H}$ is the DW Hamiltonian operator. For the free scalar field $y$ 
\beq 
 \what{H} = -\frac{1}{2}\hbar^2\ka^2 
 \frac{\der^2}{\der y^2} + 
\frac{1}{2} \frac{m^2}{\hbar^2} y^2.  
\eeq 
It corresponds to the harmonic oscillator along the field dimension $y$ 
and its spectrum $\ka m (N+\frac{1}{2})$ means that free particles of 
mass $m$ correspond to the transitions between nearby eigenstates of 
DW Hamiltonian operator.

The question arises, how the description in terms of the wave function on the finite dimensional space, $\Psi(y,x)$,  is related to standard QFT, 
e.g. the description in terms of the Schr\"odinger wave functional 
$\BPsi$$ ([y(\bx)],t)$ 
 on the infinite dimensional space of field configurations $y(\bx)$,
 which is derived from 
 canonical quantization? 
If $\Psi(y,x)$ has a probabilistic meaning of the probability amplitude of observing the value of a field $y$ at the space-time point $x$,   
then the wave functional $\BPsi$$ ([y(\bx)],t)$ should be a composition of amplitudes  
$\Psi(y,x)$ taken along the surface $\Sigma: y=y(\bx)$ at the time $t$. 
Althought the idea has been around already since 1998 \cite{myIJTP98},  
the exact form of this composition in the case of scalar field theory 
was established only recently  in 
\cite{my1201}:  
\beq 
\mbox{$\BPsi$} = 
\mbox{\sf Tr} \left .\left\lbrace \prod_\bx 
 e^{-iy(\bx)\ga{}^i\der_iy(\bx)/\ka} 
 \Psi_\Sigma (y(\bx), \bx, t)
\right\rbrace{\!\!}\right\rvert_{{\gamma}^0\ka \rightarrow \delta(\mbox{\footnotesize\bf 0})}. 
\eeq 
Here, starting from the solution of (\ref{nse}) restricted to the surface 
$\Sigma$ we construct the functional $\BPsi$ which, 
under the map  $\gamma^0\ka \rightarrow \delta(\mbox{\bf 0})$ 
(which is actually the inverse of the Clifford-algebraic "quantization map" 
from the exterior forms to Clifford numbers  in the limit 
 of vanishing "elementary volume" 
$1/\ka \rightarrow 0$), 
satisfies the canonical Schr\"odinger equation in functional derivatives.  
In this way, the standard QFT based on canonical quantization  
appears as a singular limit of QFT based on precanonical quantization.

\section{ DW formulation of vielbein gravity} 

Our earlier application of precanonical quantization in quantum gravity 
\cite{myqg}
was based on the DW formulation in metric variables. 
The resulting formulation is necessarily a hybrid 
(quantum-classical) theory 
because a part of the spin connection term in the curved space-time Dirac operator in (\ref{nse}) cannot be expressed in terms of the variables of 
 metric formulation and, therefore, quantized. 

Here we explore an alternative approach proceeding 
from the 
Lagrangian density 
written in terms of the vielbein $e^\mu_I$ and 
torsion-free spin connection variables $\omega_\alpha^{IJ},$ 
viz.  
\beq \label{lagr}
{\mathfrak L}= \frac{1}{\kappa_E} {\mathfrak e} e^{[\alpha}_I e^{\beta ]}_J (\der_\alpha \omega_\beta{}^{IJ} +\omega_\alpha {}^{IK}\omega_{\beta K}{}^J) + \frac{1}{\kappa_E}\Lambda {\mathfrak e}, 
\eeq
where $\kappa_E:= 8\pi G$ 
and ${\mathfrak e}:= \det{||e_\mu^I}||$.  
 
 If we treat the components of vielbeins  and spin connections 
as independent dynamical variables
(c.f. \cite{esposito}),  
then the corresponding polymomenta 
\beq \label{constr}
{\mathfrak p}{}^\alpha_{e^I_\beta}=
\frac{\der {\mathfrak L} }{\der (\der_\alpha e^I_\beta)}\approx 0, \quad
{\mathfrak p}{}^\alpha_{\omega_\beta^{IJ}} =\frac{\der {\mathfrak L} }{\der (\der_\alpha{\omega_\beta^{IJ})}}\approx \ \mbox{$\ \mbox{$\frac{1}{\kappa_E}$}$}
{\mathfrak e} e^{[\alpha}_Ie^{\beta ]}_{J }
\eeq
lead to what can be called the primary constraints in DW formalism, i.e. the underlying DW Legendre transformation is singular 
and there is  no unique expression of space-time 
gradients of fields in terms of the field variables and  polymomenta. 

Unfortunately, the theory of singular DW Hamiltonian systems is not yet sufficiently developed for the purposes of quantization 
(c.f. \cite{dw-constr}). We, therefore, will 
be guided by our treatment in \cite{my-dirac}.  
Namely, using the primary constraints we write down an extended DW Hamiltonian \vspace{+2pt}density 
\begin{dmath}
{\mathfrak H} = 
- \frac{1}{\kappa_E}
{\mathfrak e} e^{[\alpha}_I e^{\beta ]}_J 
\omega_\alpha {}^{IK}\omega_{\beta K}{}^J  
  - \frac{1}{\kappa_E}\Lambda {\mathfrak e}
  + \mu\cdot{\mathfrak p}{}_e
  + \lambda\cdot({\mathfrak p}{}_\omega 
  - \frac{1}{\kappa_E}{\mathfrak e} e\wedge e)~, 
\end{dmath} 
\vspace{+2pt}where 
$\mu$ and $\lambda$  are the Lagrange multipliers 
(for the sake of brevity we omit, when appropriate,  
 writing the indices explicitly). 
 The DW Hamiltonian equations given by ${\mathfrak H}$: 
\beqa
\der_{[\alpha} \omega_{\beta]}^{IJ} &\!\!\!=\lambda^{IJ}_{\alpha \beta},
\quad  
\der_\alpha {\mathfrak p}{}^\alpha_{e^I_\beta} 
&\!\!= -\frac{\der{\mathfrak H}}{\der e^I_\beta}, 
 \label{dwp}  \\
\label{dwe} 
\der_\alpha e^I_\beta &\!\!\,=\mu^I_{\alpha \beta}, 
 \quad 
\der_\alpha  {\mathfrak p}{}^\alpha_{\omega_\beta^{IJ}}
 &\!\!= - \frac{\der{\mathfrak H}}{\der \omega^{IJ}_\beta},   
\eeqa
 yield, respectively, the Einstein equations as a consequence of preservation 
 of the constraint ${\mathfrak p}^\alpha_e \approx 0$  
 and the  covariant constancy condition 
$\nabla_{}{}_\beta({\mathfrak e} e^{[\alpha}_I e^{\beta ]}_{J })=0,$  
which can be transformed into the expression of spin connection in terms of 
vielbeins and their derivatives.

 Note that eqs. (\ref{dwp}), (\ref{dwe})
  are tantamount to the preservation of 
$(n-1)$-forms constructed from the constraints (\ref{constr}), 
see  \cite{my-dirac}. 
 As the Poisson-Gerstenhaber brackets of those forms\footnote{The Poisson-Gerstenhaber (PG) 
brackets of forms have been found within  the DW Hamiltonian formalism 
in our earlier work, see e.g. [1b], and their quantization underlies the 
precanonical quantization program, see [3,4]. An attempt to construct their  Dirac-like generalization in singular DW theories is undertaken in 
\cite{my-dirac}. However, a more natural approach, 
 which is still a work in progress, would be  
 a construction of brackets from a  proper restriction of the polysymplectic 
 form to the surface of constraints 
 instead of a formal generalization of 
 the Dirac formula. 
}  
%
do not weakly vanish, the constraints in (\ref{constr}) are second class \cite{my-dirac}. 
To cope with this 
 we use our generalization of the Dirac bracket 
 to the singular DW Hamiltonian formalism \cite{my-dirac} 
 and notice that the following brackets 
 can be obtained formally without knowing the explicit form of the 
 inverse matrix of PG 
 brackets of constraints: 
\beqa \label{dbr11}
{}&\pbr{{\mathfrak p}^\alpha_e \varpi_\alpha}{e'}^D=0~,  \\
 {}&\pbr{{\mathfrak p}^\alpha_\omega \varpi_\alpha}{\omega'}^D 
 =\pbr{{\mathfrak p}^\alpha_\omega \varpi_\alpha}{\omega'} \,\,=\,\, \delta_\omega^{\omega'},
\label{dbr12} 
\\ 
{}&\!\!\hspace*{-10pt}\pbr{{\mathfrak p}^\alpha_e \varpi_\alpha}{ {\mathfrak p}_\omega}^D 
\!=\pbr{{\mathfrak p}^\alpha_e \varpi_\alpha}{\omega}^D 
\!=\pbr{{\mathfrak p}^\alpha_\omega\varpi_\alpha}{e}^D \!=0. \label{dbr13}
\eeqa 
\vspace{+2pt}
We assume that these brackets are the fundamental ones 
 whose quantization allows us to construct all other operators 
 of the theory by composition.

\section{Quantization} 
Quantization of formal Dirac brackets (\ref{dbr11}), (\ref{dbr13}) and the constraint $p_e\approx 0$ allow us to set $\what{p}_e =0$. 
It means that our precanonical wave functions 
will depend only on the spin connection and space-time variables, 
not on the vielbein variables: $\Psi=\Psi(\omega,x)$. 

{Since~the~bracket~in~(\ref{dbr12})~coincides~with~the~\mbox{familiar} 
brackets of polymomenta and field variables 
 in DW Hamiltonian formalism,  
we can use the curved space-time generalization of the operator representation of polymomenta in precanonical quantization, viz. 
 \beq \label{ptheta}
\hat{{\mathfrak p}}{}^{\alpha}_{\omega_\beta^{IJ}} 
 = -i \hbar\ka\ 
\mbox{\raisebox{-2pt}{$\vdots$}}
\hat{\mathfrak e} \hat{\gamma}{}^{[ \alpha} 
\frac{\der}{\der \omega_{\beta]}^{IJ}} 
\mbox{\raisebox{-2pt}{$\vdots$}}~,  
\eeq 
where $\hat{\mathfrak e}$ and $\hat{\gamma}{}^{ \alpha}$ are yet unknown operators and \mbox{\raisebox{-2pt}{$\vdots$}} signifies 
a potential operator ordering ambiguity.

Now, if we contract the second constraint in (\ref{constr}) with the flat  $\gamma^{IJ}$-s:\footnote{We define 
$\gamma^\nu := e^\nu_I \gamma^I, \,\, \gamma^I \gamma^J + \gamma^J\gamma^I = 
2 \eta^{IJ}$; $\eta^{IJ}$ is 
a fiducial flat Minkowskian  metric with the signature $+--...-$.}
\beq \label{c1}
{\mathfrak e} e^{[\alpha}_I e^{\beta ]}_J \gamma^{IJ} = 
{\mathfrak e} \gamma^{\alpha\beta} \approx \kappa_E {\mathfrak p}  {}^\alpha_{\omega_\beta^{IJ}}\gamma^{IJ}~, 
\eeq 
and insert the precanonical operator representation of 
$\hat{{\mathfrak p}}{}^{\alpha}_\omega$ 
in eq. (\ref{ptheta}) 
 into  the operator version of the constraint (\ref{c1}), 
 we obtain the operator representation of the curved space-time 
 Dirac matrices and 
 vielbeins, viz.   
\beq \label{gamma} 
 \what{\gamma}{}^\beta = -i \hbar\ka\kappa_E \gamma^{IJ}\frac{\der}{\der \omega_{\beta}^{IJ}}~, 
\;\; \hat{e}{}^\beta_I = -i \hbar\ka\kappa_E \gamma^{J}\frac{\der}{\der \omega_{\beta}^{IJ}}~. 
\eeq
This allows us to 
construct the DW Hamiltonian operator $\hat{H}$, 
 such that $\what{{\mathfrak H}}|_C =: \what{{\mathfrak e} H}$, 
 where  $|_C$ denotes the restriction to the surface of constraints 
 (\ref{constr}): 
\begin{equation} \label{dwh}
\what{H} = \hbar{}^2\ka^2\kappa_E \gamma^{IJ} 
\mbox{\raisebox{-2pt}{$\vdots$}} 
\frac{\der}{\der \omega_{[\alpha}^{IJ}} \frac{\der}{\der \omega_{\beta]}^{KL}} 
\omega_{\alpha}{}^{KM}\omega_{\beta  M}{}^L 
\mbox{\raisebox{-2pt}{$\vdots$}}  
- \frac{1}{\kappa_E} \Lambda~.
\end{equation}

\section{Covariant Schr\"odinger equation for quantum gravity}
The precanonical analogue of the Schr\"odinger equation for quantum gravity, 
which generalizes eq. (3),  will have the form 
\beq
i \hbar\ka 
\what{\slashed\nabla}  \Psi = 
\what{H} \hspace*{-0.0em} \Psi~,   
\eeq
were
$
\what{\slashed\nabla} 
:= 
({\gamma^\mu(\der_\mu+\omega_\mu)})^{op} 
 {\rm ~ with ~ } \omega_\mu := \frac{1}{4} \omega_{\mu IJ} \gamma^{IJ}$ 
~ is the "quantized Dirac operator". Using the operator representation 
of the curved space gamma-matrices in (\ref{gamma}) we obtain: 
\beq 
\what{\slashed\nabla}  = 
-i \hbar\ka\kappa_E \gamma^{IJ} 
\mbox{\raisebox{-2pt}{$\vdots$}} 
\frac{\der}{\der \omega_{\mu}^{IJ}} 
\left(\der_\mu +  \frac{1}{4} \omega_{\mu KL}\gamma^{KL}\right) 
\mbox{\raisebox{-2pt}{$\vdots$}}~.
\eeq
Hence,  our precanonical counterpart of the 
 Schr\"odinger equation for quantum gravity takes  the form 
\begin{dmath} \label{wdw}
\gamma^{IJ} 
\mbox{\raisebox{-2pt}{$\vdots$}} 
\frac{\der}{\der \omega_{\mu}^{IJ}} 
 \left( \der_\mu +   \frac{1}{4} \omega_{\mu KL}\gamma^{KL} 
  - \frac{\der}{\der \omega_{\beta}^{KL}}
  \omega_{[\mu }^{KM}\omega_{\beta ]M}{}^{L}  \right)
  \mbox{\raisebox{-2pt}{$\vdots$}}
   \Psi 
       + \,  \frac{\Lambda}{\hbar^2\ka^2\kappa_E^2} \Psi  = 0  .
\end{dmath} 
This equation determines the wave function $\Psi(\omega,x)$ or the transition amplitudes $\langle \omega,x |\omega',x' \rangle$ 
which provide a quantum description of geometry 
generalizing the classical 
differential geometry which uses smooth connection fields $\omega(x)$. 
This description is different from the already existing 
approaches to the quantum geometry of space-time   
in quantum geometrodynamics, loop quantum gravity or non-commutative 
geometry. 


\section{Defining the Hilbert space}
In order to specify the Hilbert space we first assume that our wave functions 
are vanishing at large $\omega$-s.    
Physically it means that the regions of very large curvature $R=d\omega + \omega\we\omega$ 
 are avoided by the wave function. The scalar product should have the form: 
$
\left\langle \Phi | \Psi \right\rangle 
:=  \int [d\omega] \Phib\Psi, 
$
where $[d\omega]$ is a Misner-like \cite{misner} covariant measure on the space of 
spin connection coefficients, which we found to have the form 
\beq 
[d\omega]={\mathfrak e}{}^{- n(n-1)}\prod_{\mu IJ} d \omega_\mu^{IJ}.
\eeq
However, since ${\mathfrak e} := \det||e_\alpha^I||$ and $e_\alpha^I$ form the inverse matrix of 
$e^\alpha_I$, which are themselves differential operators according to  (\ref{gamma}), 
 this measure is operator-valued with 
 \beq
\hat{\mathfrak{e}}{}^{-1} = 
\mbox{$\frac{1}{n!}$} \epsilon^{I_1...I_n}\epsilon_{\mu_1...\mu_n} 
\hat{e}{}^{\mu_1}_{I_1} ... \hat{e}{}^{\mu_n}_{I_n} . 
\eeq 
Hence, our scalar product has the form 
\beq 
\left\langle \Phi | \Psi \right\rangle 
:=  \int \Phib \, \what{[d\omega]}_{} \Psi.
\eeq
In order to ensure that the DW Hamiltonian operator $\what{H}$ is 
self-adjoint with respect  to this scalar product 
the expectation values are defined as follows, 
\beq \label{weo}
\langle \hat{H}\rangle  := \int \Psib \, \left(\what{[d\omega]}\hat{H}\right)_{W} \Psi,
\eeq
where the subscript $W$ denotes the Weyl ordering. 

There is a remaining local coordinate freedom in 
spin connection coefficients which should be fixed. 
A possible choice could be the 
De Donder-Fock 
 condition, 
i.e. the choice of harmonic coordinates on the average: 
\beq \label{ddf}
\der_\mu \left\langle \Psi(\omega,x) \left| \what{\mathfrak{e} g^{\mu\nu}}
\right|\Psi(\omega,x) \right\rangle = 0,  
\eeq 
where 
\beq
\what{g^{\mu\nu}} = -\hbar^2 \ka^2\kappa_E^2 
\eta^{IJ}\eta^{KL}\frac{\der^2}{\der\omega^{IK}_\mu \der\omega^{JL}_\nu}
\eeq 
is the metric operator obtained from (\ref{gamma}), 
and the operator ordering in (\ref{ddf}) is fixed by the Weyl 
ordering prescription in (\ref{weo}).
Note that in the present formulation the coordinate condition is 
imposed on the wave functions $\Psi(\omega,x)$ rather than on the 
metric or vielbein fields. 

\section{Precanonical quantum cosmology, a toy model}
Let us consider $n=4$,  $k=0$ FLRW metric with a  harmonic 
time coordinate $\tau$
\beq
ds^2=a(\tau)^6d\tau^2 - a(\tau)^2 d\bx^2 
= \eta_{IJ}e^I_\mu e^J_\nu dx^\mu dx^\nu .
\eeq
Then $e^0_\nu = a^3 \delta^0_\nu$ and  $e^J_\nu = a \delta^J_\nu$ 
for $J=1,2,3$,   
and the non-vanish\-ing components of spin connection are 
$\omega_i^{0I}= - \omega_i^{I0}
= {\dot{a}}/{2a^3} =:\omega$, where $i=I=1,2,3.$



In this simple case 
 there is 
only the $\Lambda$-term which remains in 
the DW Hamiltonian operator, eq.~(\ref{dwh}),   
and eq.~(\ref{wdw}) takes the form 
\beq
\Big ( 2 \sum_{i=I=1}^3 
 \alpha^I\der_\omega\der_i + 
 3\omega\der_\omega + 
 \lambda \Big )\Psi=0,  
\eeq
 where  $\alpha^I:= \gamma^{0I}$ and 
$\lambda:= \frac{3}{2} + {\Lambda}/({\hslash\ka\kappa_E})^2$,  
if the Weyl ordering is used. Note that the correct value of 
$\Lambda$ can be obtained from the constant of order  unity  
which results from the operator ordering, provided  $\ka \sim 10^{-3}GeV^3$.

By separating variables $\Psi := u(x)f(\omega)$ we obtain the 
equation on $u$: 
$$2 \sum_{i=I} \alpha^I \der_i u = iq u,$$ 
where the imaginary unit  comes from the anti-hermicity of $\der_i$, 
and the  equation on the wave function in $\omega$-space: 
$$(iq\der_\omega + 3\omega\der_\omega + \lambda ) f=0,$$
whose solution 
 $f\sim (iq+3\omega)^{-\lambda}$ 
yields the probability density  
(similar to $\mathsf{t}$-distribution) 
\beq
\rho(\omega) := \bar{f}f \sim (9\omega^2 + q^2)^{-\lambda}.  
\eeq 
At  $\lambda > 1/2$,  which is required by 
{\sf L}${}^2[(-\infty,\infty),[d\omega]=d\omega]$ 
normalizability in $\omega$-space, this distribution   
has a bell-like  shape  centered at the  zero 
expansion rate $\dot{a}=0$. 
The 
 most probable expansion 
 rate can be shifted by accepting complex 
values of $q$,  and the inclusion of minimally coupled matter fields 
changes $\lambda$.

Although our toy  model bears some similarity with the minisuperspace models, 
 %
its origin and the content are different. It is obtained from the full quantum Schr\"odinger equation 
 (\ref{wdw}) 
when the field $\omega$ is one-component, 
rather than 
 via quantization of 
 a 
 reduced mechanical model deduced under 
the assumption of spatial homogeneity.  
In fact, the naive assumption of spatial homogeneity of the wave 
function: $\der_i\Psi =0$, or $q=0$, would not be compatible with its normalizability  in $\omega$-space.  
Instead, our model 
 implies a quantum gravitational structure of space 
at the scales $\sim\!\mathsf{Re}$\mbox{$\frac{1}{q}$}  
and $\sim\!\mathsf{Im}$\mbox{$\frac{1}{q}$} given by $u(x)$.  


\section{Discussion}
This paper presents rather a prototype of the theory of quantum gravity 
which results from a brute force implementation of precanonical quantization in general relativity. 
Our goal  was to highlight the potential of the approach in constructing 
a mathematically well-defined, nonperturbative, covariant and background independent formulation of quantum gravity and 
 to discuss some features of the resulting theory.  

 A difficulty  we faced is related to the second class constraints 
in the DW Hamilton\-ian formulation whose complete analysis is 
not well understood. 
In fact, in addition to the brackets  
in (\ref{dbr11})--(\ref{dbr13}),  one can also calculate 
e.g. 
$\pbr{e_I^\alpha\varpi_\alpha}{\omega_\nu^{IJ}}^D 
$ 
which, however, is hardly possible to quantize 
directly using the Dirac's rule, because it 
explicitly depends on the generalized inverse of 
the 
 matrix of 
 PG brackets of constraints. 
Because of this we proceeded from 
 quantization of brackets 
in (\ref{dbr11})--(\ref{dbr13}) alone, hoping that the results can be compatible with the right hand side of the commutators similar to 
 $[\what{e_I^\alpha\varpi_\alpha},\what{\omega}{}_\nu^{IJ}]$ 
when $\what{e_I^\alpha\varpi_\alpha}$ is treated as a composite operator. 
However, the validity of this workaround 
and formal calculation of Dirac brackets (\ref{dbr11})-(\ref{dbr13}) 
needs further scrutiny. 

The immediate consequences of our approach are that 
(i) the quantum dynamics is described in the space of spin connection coefficients $\omega$, (ii) the metric becomes a composite operator, eq. (26), (iii) the quantum description of geometry is achieved in terms 
of the transition amplitudes $\langle \omega,x |\omega',x' \rangle$ which obey the covariant Schr\"odinger equation, eq. (20). 

We also noticed that the correct value of the cosmological constant can be obtained from the dimensionless number of order unity, which appears from the ordering of operators in eq. (\ref{wdw}), if the parameter $\ka$ of 
 precanonical quantization corresponds to the energy scale of roughly 
 $10^2 MeV$,  which is a rather unexpected fact to be understood. 
 
One of the motivations to consider  the vielbein formulation as a starting 
point of precanonical quantization was a desire to understand if the hybrid formulation of precanonical quantum gravity in  \cite{myqg}, 
with its ``bootstrap condition" and the space-time emerging 
from the self-consistency requirement,  
is indeed a feature of precanonical quantum 
gravity or a reflection of limitations of the metric formulation.  
Whereas the present consideration seems to suggest the latter, the certain answer 
 requires clarity regarding the treatment of the second class 
 constraints in DW formulation of vielbein gravity.


\begin{thebibliography}{99}
\bibitem{dw} 
H.~Kastrup, 
Canonical theories of dynamical systems in physics, 
{\em Phys. Rep.} {\bf 101} (1983) 1-167;  \\ 
I.V.~Kanatchikov, 
Canonical structure of classical field theory in~the~poly\-momentum phase space,
{\em Rep. Math. Phys.} {\bf 41} (1998) 49-90, 
 {\tt hep-th/9709229}. 

\bibitem{garcia}
M.~Gotay, J.~Isenberg, J.~Marsden,  
Momentum maps and classical relativistic fields. Part 1: Covariant field theory,
{\tt physics/9801019};\\
 Momentum maps and classical relativistic fields. 
Part II: Canonical analysis of field theories, 
{\tt math-ph/0411032}.
 
\bibitem{myprecanonical} 
I.V.~Kanatchikov, 
On quantization of field theories in poly\-momentum variables, 
in {\em AIP Conf. Proc.} {\bf 453} (1998) 356-367, 
{\tt hep-th/9811016};  \\
I.V.~Kanatchikov, De Donder-Weyl theory and a hypercomplex extension of quantum~mechanics 
to field theory, 
{\em Rep. Math. Phys.} {\bf 43} (1999) 157-170, {\tt hep-th/9810165};\\
I.V.~Kanatchikov, Geometric (pre)quantization in the polysymplectic approach to field theory,
{\tt hep-th/0112263}. 


\bibitem{myIJTP98}
I.V.~Kanatchikov,  
 Towards the Born-Weyl quantization of fields, 
{\em Int. J. Theor. Phys.} {\bf 37} 333-342 (1998), {\tt quant-ph/9712058};\\ 
 I.V.~Kanatchikov, 
 Precanonical quantization and the Schr\"o\-dinger wave functional,
{\em Phys. Lett.} {\bf A283} 25-36 (2001), {\tt hep-th/0012084}. 

\bibitem{my1201} I.V.~Kanatchikov, 
Precanonical quantization and the Schr\"o\-dinger~wave functional 
revisited, 
{\tt arXiv:1112.5801}. 

\bibitem{myqg} 
I.V.~Kanatchikov, 
Precanonical quantum gravity:~quantization without the space-time decomposition, 
{\rm Int. J. Theor. Phys.} {\bf 40}  1121-1149 (2001),
{\tt gr-qc/0012074}. \\
See also: I.V.~Kanatchikov, 
From the De Donder-Weyl Hamiltonian formalism to quantization of gravity,
{\tt gr-qc/9810076}; 
Quantization of gravity: yet another way, 
{\tt gr-qc/9912094}; 
Precanonical perspective in quantum gravity, 
{\em Nucl. Phys. Proc. Suppl.} {\bf 88}  326-330 (2000),
{\tt gr-qc/0004066}.

\bibitem{esposito} 
G.~Esposito, G.~Gionti, C.~Stornaiolo, 
 Space-time covariant form of Ashtekar's constraints, 
{\em Nuovo Cim.} {\bf 110B} 1137-52 (1995), 
{\tt gr-qc/9506008}.

\bibitem{dw-constr}
X.~Gracia, R.~Martin, N.~Rom\'an-Roy,
Constraint algorithm for k-presymplectic Hamiltonian~systems.
Application to singular field theories,
	{\tt arXiv:0903.1791}; \\
M.~Castrill\'on L\'opez, J.E.~Marsden, 
Some remarks on Lagrangian and Poisson reduction for field theories, 
\emph{J. Geom. Phys.} {\bf 48} 52-83 (2003);\\
G.~Sardanashvily,
Constraints in polysymplectic (covariant) Hamiltonian formalism, 
{\tt math-ph/0008024}; \\
 C.M.~Campos, M.~de~Leon, D.M.~de~Diego,
 Constrained variational calculus for higher order classical field theories,
 {\tt arXiv:1005.2152}. 
	

\bibitem{my-dirac}
I.V.~Kanatchikov,
On a generalization of the Dirac bracket in the De Donder-Weyl 
Hamiltonian formalism, 
In: {\sl Differential Geometry and its Applications, } Proc. 10th Int. Conf. on Diff. Geom. \& Appl., Olomouc, August 2007, O. Kowalski, D. Krupka, O. Krupkov\'a and J. Slov\'ak (Eds.) (World Scientific, Singapore, 2008) 615-625, 
{\tt arXiv:0807.3127}.

\bibitem{misner}	
C.~Misner, 
Feynman quantization of General Relativity,  
{\em Rev. Mod. Phys.} {\bf 29} 497-509 (1957). 

\end{thebibliography}
\end{document}